\documentclass{elsart}
\usepackage{graphicx}
\usepackage{amsmath}
\usepackage{amssymb}

\newcommand{\het}    {\mbox{$ ^{3}{\mathrm{He}} $}~}

\def\NIMA#1#2#3{{\rm Nucl.~Instr.~and~Meth.} {\bf{A#1}} (#2) #3}

\def\PLB{{\em Phys. Lett.}  B}

\def\PRD#1#2#3{{\rm Phys. Rev.} {\bf{D#1}} (#2) #3}

\def\PRL#1#2#3{{\rm Phys.~Rev.~Lett.} {\bf{#1}} (#2) #3}
\def\PLB#1#2#3{{\rm Phys.~Lett.} {\bf{B#1}} (#2) #3}

\begin{document}

\begin{frontmatter}

\title{Bolometric calibration of a superfluid $^3$He detector for Dark Matter search:
direct measurement of the scintillated energy fraction for neutron, electron and muon
events}

\author[crtbt]{C. B. Winkelmann},
\author[crtbt]{J. Elbs},
\author[crtbt]{Yu. M. Bunkov\thanksref{corr}},
\author[crtbt]{E. Collin},
\author[crtbt]{H. Godfrin},
\author[hut]{and M. Krusius}
\thanks[corr]{Corresponding author : yuriy.bunkov@grenoble.cnrs.fr (phone:
+33 4-76 88 12 52)}
\address[crtbt]{Centre de Recherches sur les Tr\`es Basses Temp\'eratures,
  CNRS, laboratoire associ\'e  l'Universit\'e Joseph Fourier et
   l'Institut Polytechnique de Grenoble, BP166,  38042, Grenoble, France}
\address[hut]{Low Temperature Laboratory, Helsinki University of Technology,
FIN-02015 HUT, Espoo, Finland}


\begin{abstract}
We report on the calibration of a superfluid $^3$He bolometer
developed for the search of non-baryonic Dark Matter. Precise
thermometry is achieved by the direct measurement of thermal
excitations using Vibrating Wire Resonators (VWRs). The
heating pulses for calibration were produced by the direct quantum process of quasiparticle
generation by other VWRs present. The bolometric
calibration factor is analyzed as a function of temperature and excitation level of
the sensing VWR. The calibration is compared to bolometric measurements of the
nuclear neutron capture reaction and heat depositions by
cosmic muons and low energy electrons. The comparison allows a
quantitative estimation of the ultra-violet scintillation rate of
irradiated helium, demonstrating the possibility of efficient electron recoil event rejection.
\end{abstract}


\end{frontmatter}

\newpage

\section{Introduction}

Superfluid $^3$He-B at ultra-low temperatures is an appealing target
material for bolometric particle detection \cite{First,Second,BOLO}, with complimentary features
to the currently most performing germanium- and
silicon-based detectors like Edelweiss \cite{edelweiss} and CDMS
\cite{cdms}. Among the attractive features of $^3$He are
the clear signature of neutron background events due to the nuclear capture reaction of
neutrons in $^3$He.  Furthermore, the huge density of unpaired neutrons leads to a significantly enhanced axial
interaction cross-section within a large class of WIMP models \cite{machetnim,machetplb}. Its single thermal energy
reservoir with the possibility of direct thermometry, the very
low experimental base temperature of about 100 $\mu$K, together with the possibility of electron recoil rejection make $^3$He a very promising target
material for Dark Matter search.

The nuclear neutron capture reaction $^3\rm{He}+n\rightarrow
^3\rm{H} + p$ releases 764 keV kinetic energy to its products, and
was detected bolometrically in the superfluid \cite{BOLO}. The
comparison of the detected neutron peak to bolometric calibrations
of the detector was then used \cite{Gren} as a test of topological
defect creation scenarios  in fast second order phase transitions
\cite{nature85}. A bolometric calibration method of the detector
was first described in \cite{Gren2}. In the following years, the detection
threshold of the bolometer could be lowered, which allowed the
identification of a broad peak at about 60 keV, that could be
attributed to cosmic muons, in agreement with numerical predictions.
In very recent measurements on a Dark Matter detector prototype, the
detection threshold reached the keV level and the low energy
electron emission spectrum from a $^{57}$Co source could be
resolved \cite{moriond,manu}. Since a neutralino interaction with a
$^3$He nucleus in the detector is expected to deposit a maximum
energy of about 5.6 keV, the current detector design therefore
already displays the required bolometric sensitivity.

The recent improvement of the sensitivity was partly made possible
by lowering the working temperature of the detector cells from about
150 to 130 $\mu$K. This temperature decrease of less than 20\%
represents a decrease in the thermal enthalpy of the cells by an
order of magnitude. On the other hand, at lower temperature the
weaker coupling of the VWR to the superfluid results nevertheless in
a greater response time of the thermometer. Parallely, the
non-linear velocity dependence of the friction with the superfluid
\cite{fisher}, is also of greater importance at the lower
temperatures and the higher VWR-response signals currently used.
This article hence proposes a generalization of the methods
described in \cite{Gren2} for a more profound and comprehensive
understanding of the bolometric detector (sections 2 to 4) and its
calibration by mechanical heater pulses (section 5). The comparison
of the bolometric calibration with detection spectra from known
energy sources like neutrons (section 6) as well as low energy
electrons and muons (section 7) allows then a precise estimation of the
ultra-violet (UV) scintillation rate of helium for several types of ionizing radiation.

\section{Principle of detection}

The current \het particle detector \cite{moriond,manu} consists
of\cite{Gren2}
three adjacent cylindrical copper cells, of volume $V=$ 0.13
cm$^3$ each, immersed inside a superfluid \het bath at ultra-low
temperature, at about 130 $\mu$K. A 200\,$\mu$m diameter orifice in
the wall of the cells connects the $^3$He inside the cell with the
surrounding heat reservoir of superfluid. An energy deposition by an
incident particle creates a cloud of quasiparticles in the cell,
which comes to internal thermal equilibrium via collisions on the
cell walls within $\sim$ 1 ms. The excess ballistic quasiparticles,
of momentum $p\approx p_F$, then leak out of the cell with a time
constant $\tau_b$ of a few seconds, which is determined mainly by
the size of the orifice .

\begin{figure}[ht]
\begin{center}
\includegraphics[scale=0.5,angle=0]{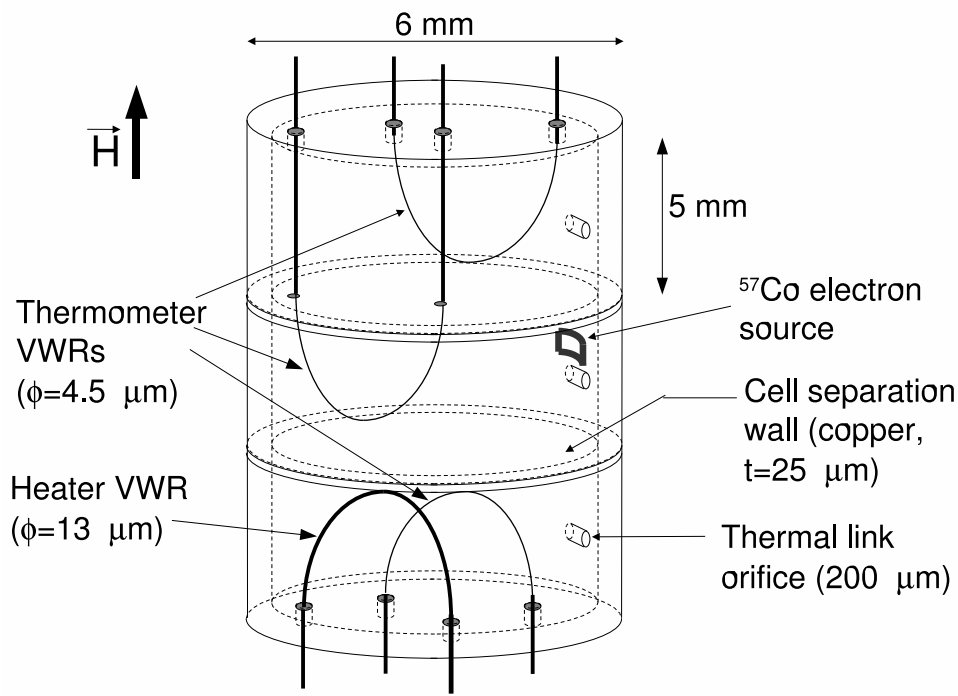}
\caption{Schematic setup of the 3-cell bolometric detector prototype. Each cell contains a VWR thermometer. The cells are in weak thermal contact with the outer $^3$He bath through the orifice. The presence of three adjacent cells allows discrimination through coincident detection. One of the cells contains the $^{57}$Co source (section 7), another one contains an extra VWR for the purpose of bolometric calibration by heater pulses (section 5).}
\label{fig:3cell}
\end{center}
\end{figure}

The Vibrating Wire Resonators (VWRs) used for thermometry are thin
(4.5 $\mu$m diameter in our setup) superconducting NbTi filaments
bent into an approximately semi-circular shape of a few mm, with
both ends firmly fixed \cite{thermo}. The VWR is driven by a Laplace
force imposed by an AC current close to its mechanical resonance
frequency ($\approx$500 Hz), and oscillates perpendicularly to its
main plane with an $rms$ velocity $v$. The motion is damped by
frictional forces, of total amplitude $F(v)$, mainly due to momentum
transfer to the quasiparticles of the surrounding superfluid.
Relatively fast thermometry is then achieved by measuring
continuously the damping coefficient $W\propto F/v\ $ of a VWR
driven at resonance. In the low oscillation velocity limit, the
friction is of a viscous type $F\propto v$. $W$ is thus velocity
independent and can be written as \cite{fisher-bis}
\begin{equation}
W_{0}=\alpha\ \exp(-\Delta/k_BT),
\label{damping}
\end{equation}
where $\Delta$ is the superfluid gap at zero
temperature. The value
of the prefactor $\alpha$ is determined by both the microscopic properties of the
liquid and the geometry of the VWR. At 0 bar and for the VWRs used in the detector,
of density $\rho=6.0$ g/cm$^3$ and radius $a=2.3 \ \mu$m, $\alpha$
is of the order of 1-2 10$^5$ Hz, depending on the exact geometrical
features. In quasi-static conditions, the low velocity damping
coefficient $W_0$ is measured as the Full Width at Half-Height
$\Delta f_2$ of the mechanical oscillator's lorentzian resonance in
frequency space.

\begin{figure}[ht]
\begin{center}
\includegraphics[scale=0.5,angle=0]{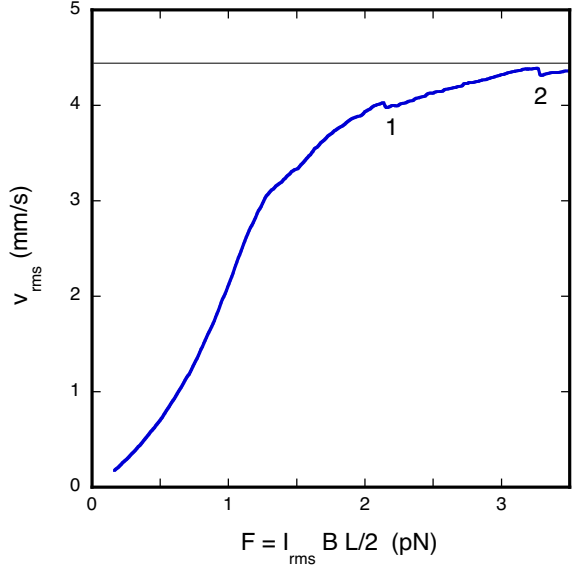}
\caption{Isothermal force-velocity ($\equiv (I,V)$) curve of a
VWR
immersed in superfluid $^3$He at ultra-low temperatures. For
$rms$ oscillation velocities below 2 mm/s, the damping is well
described by (\ref{eq:non-lin-damping}). The decrease of the damping
coefficient ($\sim$ inverse slope) at higher velocities actually
leads to an increase of the calibration factor (section 4). At
velocities above $\sim$2.5 mm/s, the damping increases again. Points
1 and 2 show two discontinuities of the force-velocity curve (see
text). The pair-breaking velocity is reached at about 4.5 mm/s.}
\label{fig:I-V}
\end{center}
\end{figure}

At finite oscillation velocities however, effects beyond the
two-fluid model
discovered in Lancaster, lead to a velocity dependent
friction which is actually decreasing at higher velocities (Fig.
\ref{fig:I-V}). A model proposed by Fisher {\em et al.}
\cite{fisher,fisher-bis} leads to a velocity dependent damping
\begin{equation}
W_{L}(v)=W_0 \times L(v/v_0),
\label{eq:non-lin-damping}
\end{equation}
where  the function
\begin{equation}
L(u)=(1-e^{-u})/u
\end{equation}
gives a very good approximation of the reduction of the
damping of
the VWR for $v$ up to a characteristic oscillation velocity
$v_0\sim k_BT/p_F\approx$ 1-2 mm/s.

At higher velocities, of about 3-4 mm/s, the local flow field
around
sharp edges on the VWR surface starts to get turbulent
\cite{bradley}. At even higher velocities, the critical velocity for
pair-breaking is achieved. Both cases correspond to a strong
enhancement of the friction, with a strongly non-linear, sometimes
even discontinuous, velocity dependence (Fig. \ref{fig:I-V}). While
the Lancaster-type non-linear coupling to the superfluid can easily
be accounted for by using equation (\ref{eq:non-lin-damping}), the
velocity has thus still to be kept small enough for both the locally
turbulent and the pair-breaking regime to be avoided.

\section{Response of the detector to a heat release}

As the quasiparticle density, the temperature, and thus $W_0$ as
defined by (\ref{damping}), increase nearly instantaneously ($\sim1$
ms) after a particle impact, the mechanical VWR only responds to
this modification over a timescale inversely proportional to the
dissipative coupling
\begin{equation}
\tau_w=\frac{1}{\pi W_L}.
\label{eq:delay}
\end{equation}
This can therefore result in non-negligible response times ($>$1 s)
for low damping coefficients, i.e. narrow resonances and low
temperatures. While recording transitory heating events, the
$dynamically$ measured damping $W_{mes}(t)$ is thus a
non-equilibrium measurement of $W(t)$.

\begin{figure}[ht]
\begin{center}
\includegraphics[scale=0.5,angle=0]{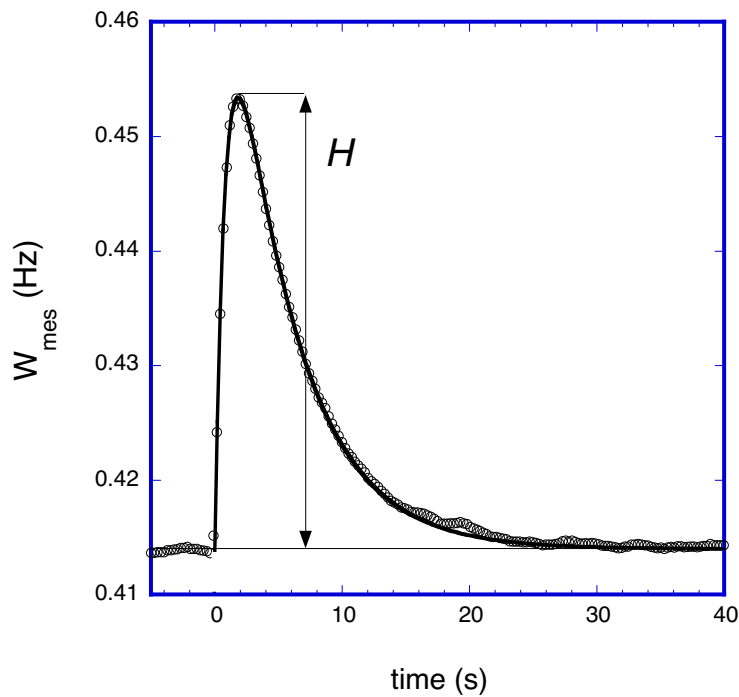}
\caption{Bolometrically recorded event ($\sim100$\,keV) (circles). The fit by
(\ref{eq:peak}) uses
the input parameters $W_{base}=414$\,mHz and
thus $\tau_w=1/\pi W_{base}=$0.77\,s. The cell relaxation time
$\tau_b=5.0$\,s is fixed for all events of the entire experimental
run. The only free parameters are the start time of the event and
and $A$. Here the fit fields $A=56.0$\,mHz, in good consistency with
$H\times G^{-1}(W_{base},\  \tau_b)=55.4$\,mHz.} \label{fig:event}
\end{center}
\end{figure}

A simple model of a sudden increase of $W$ from its baseline
$W_{base}$ by an amount $A\ll W_{base}$ at $t=0$, and the subsequent
exponential relaxation of the quasiparticle excess via the orifice
\begin{eqnarray}
W(t)=W_{base} + A \  e^{  - t/\tau_{b} } \   \Theta(t),
\label{eq:event}
\end{eqnarray}

leads to a dynamic damping measurement given by
\begin{eqnarray}
W_{mes}(t)=W_{base} + A \frac{\tau_{b}}{ \tau_{b}-\tau_{w}}  \left[  e^{  - t/\tau_{b} }           - e^{- t /\tau_{w}} \right]         \Theta(t).
\label{eq:peak}
\end{eqnarray}
$\Theta(t)$ is here the Heaviside function accounting for the instantaneous heat input and since $A$ is a small
perturbation, $\tau_w\approx constant=1/\pi W_{base}$ is assumed.
The good agreement of (\ref{eq:peak}) with experimentally detected events can be seen
on figure \ref{fig:event}. The maximum geometrical height
$H=|W_{mes}-W_{base}|_{max}$ of the response peak is then related to
$A$ by the function
\begin{eqnarray}
G(W_{base}, \tau_{b})=H/A= \left( \tau_{w}/\tau_{b} \right)
  ^{     \frac{\tau_{w}}{\tau_{b}-\tau_{w}}   } .
\label{eq:HoverA}
\end{eqnarray}
For say $W_{base}=0.414$ Hz ($T=129\ \mu$K) and $\tau_b=5.0$ s, the
slowing down of the VWR response is hence responsible for the loss
of 29 \% of the signal amplitude.

\section{Calibration factor}

Since both the enthalpy and $W_0$ are roughly proportional to the
quasiparticle density, the bolometric calibration factor
$\sigma$ of a given cell, defined as the conversion factor relating the
amplitude $A$ ($\ll W_0$) of the rise in $W_0$ in (\ref{eq:event})
to the heat $U$ released
\begin{eqnarray}
\sigma =  A/U,
\label{eq:sens_def}
\end{eqnarray}
is a slow function of temperature. Around 130 $\mu$K, the temperature dependence of the specific heat
of $^3$He-B  is given to a good approximation by
\cite{Gren2,vollhardt}
\begin{eqnarray}
C =C_0 \ (T_c/T)^{3/2} \ \exp(-\Delta/k_BT),
\label{numheatcapacity-bis}
\end{eqnarray}
where $T_c$ is the superfluid transition temperature and $C_0=
1.7$ mJ/Kcm$^3$. Writing $\delta U=CV\delta T$ allows to express the
calibration factor using (\ref{damping}), (\ref{eq:sens_def}) and
(\ref{numheatcapacity-bis}) in the low velocity limit as
\begin{eqnarray}
\sigma_0  (T) =\frac{ \alpha\Delta}{k_BC_0VT_c^{3/2}}\  \frac{1}{\sqrt{T}}.
\label{eq:sigma0}
\end{eqnarray}
Taking into account the velocity dependence of the damping
(\ref{eq:non-lin-damping}) leads to an enhancement of the
calibration factor at higher velocities over $\sigma_0$ by a factor
\begin{eqnarray}
f(u, t') =\sigma(v)/\sigma_0= t' \left[  \left( 1+t'^{-1} \right)  \frac{\sinh(u/2)}{u/2} -e^{-u/2} \right] \frac{\sinh(u/2)}{u/2},
\label{eq:non-lin-calib-function}
\end{eqnarray}
where $t'=k_BT/\Delta$ and $u=v/v_0$. At velocities $v\approx v_0$,
this results consequently in an enhancement of the calibration
factor by about 12 \%. This analysis is valid for velocities up to
$\sim1.5\ v_0$, before the other dissipative mechanisms discussed
set on.

Adding in (\ref{eq:HoverA}) leads thus to the conclusion that the
relation of the geometric height $H$ of a peak following a heat
deposition $U$ depends on both the baseline $W_{base}$ and the $rms$
velocity $v$ of the oscillation following
\begin{eqnarray}
H=\sigma_0 \ f(u, t') \times G(W_{base}, \tau_b)\times U.
\label{eq:full-line}
\end{eqnarray}

\section{Bolometric calibration by heater pulses}

At ultra-low temperatures, simulation of heating events by Joule
heating is inefficient in superfluid $^3$He because of the diverging
thermal (Kapitza) resistance at the solid-liquid boundaries. Bradley
{\em et al.} \cite{BOLO} proposed a heating method of the
superfluid, based on the mechanical friction of the oscillating VWR
with the liquid. Energy is injected through a second VWR present
(the "heater" VWR, as in the lowest of the 3 cells in Fig.
\ref{fig:3cell}), by driving it over a short time with an excitation
current at its resonance frequency. The energy is then dissipated to
the superfluid via the velocity dependent frictional coupling. One
should bare in mind that the electrically injected power is firstly
transformed into mechanical energy of the resonator before being
transfered to the liquid by friction. Therefore, even for a short
excitation pulse of the heater ($\delta t\sim 100$ ms), the heat
release to the liquid nevertheless takes place on a characteristic
timescale given by (\ref{eq:delay}), where the damping $W$ is
relative to the heater wire.

\begin{figure}[ht]
\begin{center}
\includegraphics[scale=0.47,angle=0]{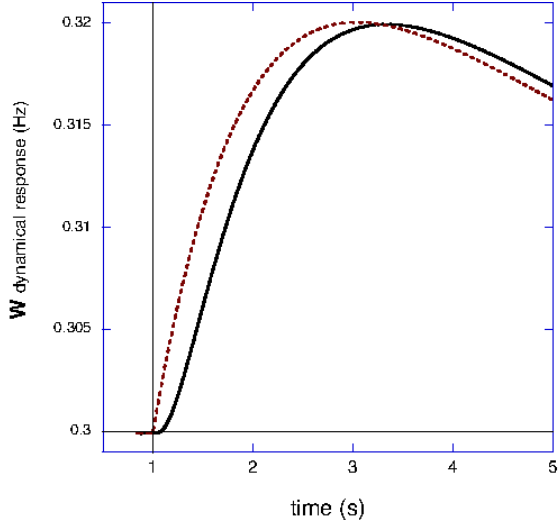}
\caption{Simulation of the response of the thermometer VWR to an instantaneous heat deposition at $t=1$ s (dotted line) and to a mechanical heater pulse of same total energy but releasing the heat slowly (continuous line) because of $\tau_w^{heater}$. While the response to the instantaneous event is well described by (\ref{eq:peak}), the response to the slow heat release has a different shape during the rise, but is then rather identical exept for a time shift due to the slower release of the energy by the heater. The height $H$ of the two peaks is nevertheless identical within 1 \% at 0.3 Hz (128 $\mu$K) and much less at higher temperatures.}
\label{fig:simu}
\end{center}
\end{figure}

The energy deposition by a heater pulse is hence not instantaneous,
even in the case of extremely short pulses. We have estimated
quantitatively this heat release, in order to verify wether a
mechanical heater pulse is indeed an acceptable 'simulation' of a
real - instantaneous - event. If one assumes velocity independent
damping of the heater resonator, its amplitude of oscillation must
increase linearly during the pulse and then relax exponentially with
timescale $\tau$. We have modeled numerically under these
assumptions the heat release to the superfluid in such a heater
pulse, as well as the response of the thermometer wire in the
bolometric cell. One result and its comparison to the response
calculated for an instantaneous energy release as say by an incident
particle, are given in Fig. \ref{fig:simu}. One can see that even
though the shape of the thermometer's response is quite different,
especially during the rise of the response, the maximum amplitude is
still the same within less than 1 \% for $W_{base}>0.3$\,Hz. A fit of the
response by (\ref{eq:peak}) is therefore not appropriate in the case
of mechanical heater pulses, while the geometrical maximum height
$H$ of the response peak is still a good parameter for comparing
pulses to instantaneous events.

A more serious issue for the bolometric calibration by mechanical
heater pulses are the intrinsic losses within the heater wire. The
intrinsic damping (i.e. in the absence of liquid) of a VWR can be
well modelled by a viscous friction term which therefore simply adds
as a constant extra term $W_{int}$ to the measured damping $W$ in
(\ref{damping}). In our experiment, we measure the heater VWR
intrinsic width as the limiting value of the damping in the low
temperature limit, to be $W_{int}^{heater}=77\pm6$ mHz. As the
thermal quasiparticle damping vanishes at low temperatures, the
intrinsic losses in the heater represent a fraction
$W_{int}^{heater}/(W_{int}^{heater}+W_{L}^{heater})$ of the injected
energy and constitute a non-negligible correction to the energy
injection through the heater wire.

\begin{figure}[ht]
\begin{center}
\includegraphics[scale=0.4,angle=0]{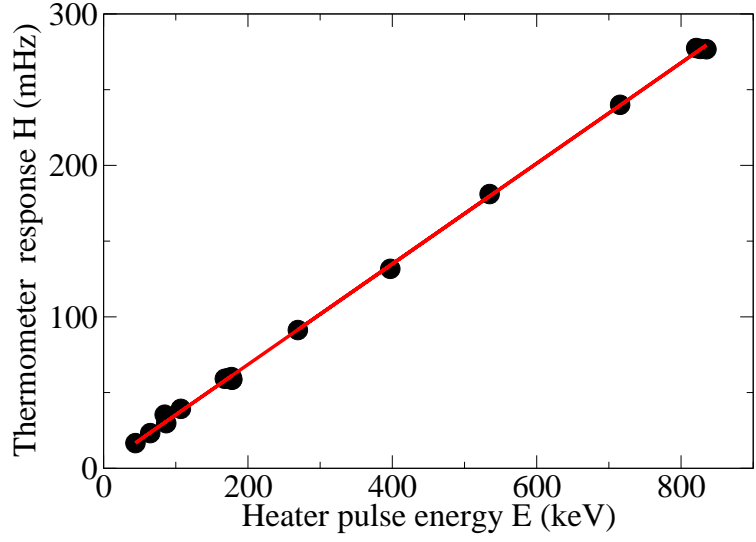}
\caption{Height $H$ of the response of the thermometer VWR to mechanical
heater pulses of energy $E$ (bullets) and linear fit (data taken at $W_{base}=1.7$ Hz).}
\label{fig:H/E}
\end{center}
\end{figure}

\begin{figure}[ht]
\begin{center}
\includegraphics[scale=0.45,angle=0]{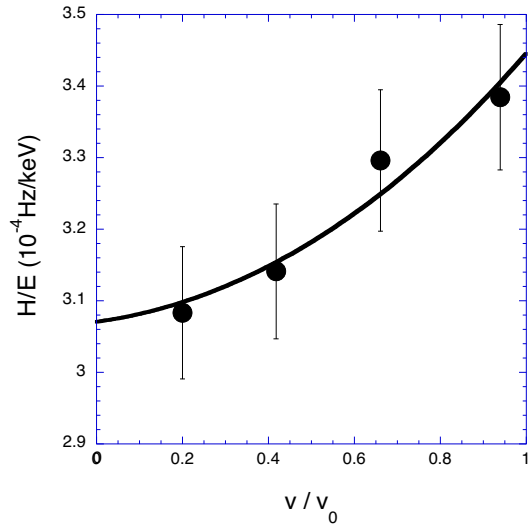}
\caption{Dependence of the observed slopes of $H(E)$ as in Fig.
\ref{fig:H/E} as a function of the reduced oscillation velocity of
the thermometer VWR ($W_{base}=4$ Hz). The continuous line represents the expected
evolution from (\ref{eq:non-lin-calib-function}), the only free
parameter being $H/E(v=0)$.} \label{fig:slope/v}
\end{center}
\end{figure}

Let us now consider the energy and velocity dependence of the
detector response. At constant oscillation velocity, the height $H$
of the response is experimentally observed to be linear in the
injected energy in the range of a few hundred keV (Fig.
\ref{fig:H/E}), which confirms the assumptions leading to
(\ref{eq:sens_def}). Furthermore, a comparison of the observed
slopes versus the VWR's oscillation velocity (Fig.
\ref{fig:slope/v}) are in good agreement with the enhancement of the
response at higher velocities, as given by
(\ref{eq:non-lin-calib-function}). This analysis allows therefore to
extract the velocity independent limit of the calibration factor
$\sigma_0$.

\begin{figure}[ht]
\begin{center}
\includegraphics[scale=0.6,angle=0]{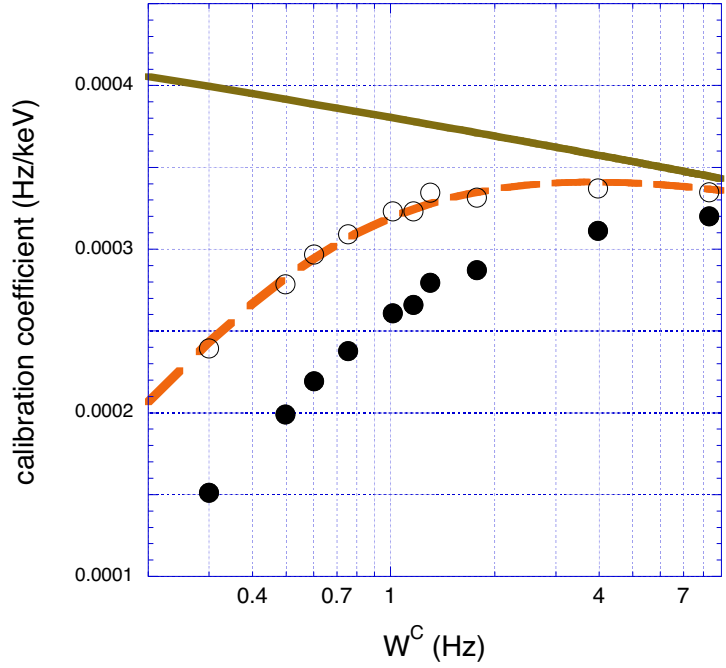}
\caption{Calibration of the observed H/E in the zero velocity limit
from mechanical heater pulses ($\bullet$). The data are then
corrected by (\ref{eq:HoverA}) to account for the slowing down of
the thermometer at low temperatures ($\circ$). Considering the fact
that a fraction $W^{heater}_{int}/W^{heater}_{tot}$ of the injected
energy is dissipated inside the heater VWR, the result can be fitted
by (\ref{eq:sigma0})(dashed line). The only free parameter is the overal
prefactor of (\ref{eq:sigma0}) which allows then to express
$\sigma_0(T)$ (continuous line).} \label{fig:calib}
\end{center}
\end{figure}

Taking into account intrinsic losses in the heater as well as the features of the thermometer VWR response deduced in the previous sections, one can fit the mechanical pulse calibration data versus baseline (i.e. versus temperature) as shown in Fig. \ref{fig:calib}. If $W_{int}^{heater}$ is left as a free parameter, the fit yields a value of $W_{int}^{heater}=74$ mHz and a calibration factor of $\sigma_0=3.8\ 10^{-4}$ Hz/keV at a baseline of 1 Hz (\ref{fig:calib}). Imposing $W_{int}^{heater}=77$ mHz leaves the overall prefactor of $\sigma_0$ (equivalent to the knowledge of $\alpha$) as the only free fitting parameter, with a value only 0.5\% different.

\clearpage
\newpage

\section{Neutrons}

The heat release to the superfluid following a nuclear neutron
capture reaction $^3$He + $n\ \rightarrow \ ^3$H + $p$ was observed
in \cite{Gren} to be at 0 bar about 15 \% less than the 764 keV
known to be released by the reaction. For a microscopic calculation of the
stopping of the neutron capture products in $^3$He  see
\cite{meyer}. A substantial part of the energy deficit can be
attributed to ultra-violet (UV) scintillation \cite{mckinsey-pra}.
The UV fraction was estimated in \cite{Gren} to account for about half of the
energy deficit, concluding that the remaining $\approx$\,60 keV were
trapped in the form of topological defects.

Neutrons are an extremely valuable tool for the characterization of
the superfluid bolometers because they provide a narrow and
cell-geometry independent peak at rather high energies. Once the
bolometric heat release of neutrons in $^3$He inambiguously
determined, no other calibration, e.g. by pulses, is necessary. The
comparison of the evolution of the observed neutron peak with
temperature also allows to test directly, without the use of
mechanical heater pulses, the assumptions leading to
(\ref{eq:full-line}).

\begin{figure}[ht]
\begin{center}
\includegraphics[scale=0.45,angle=0]{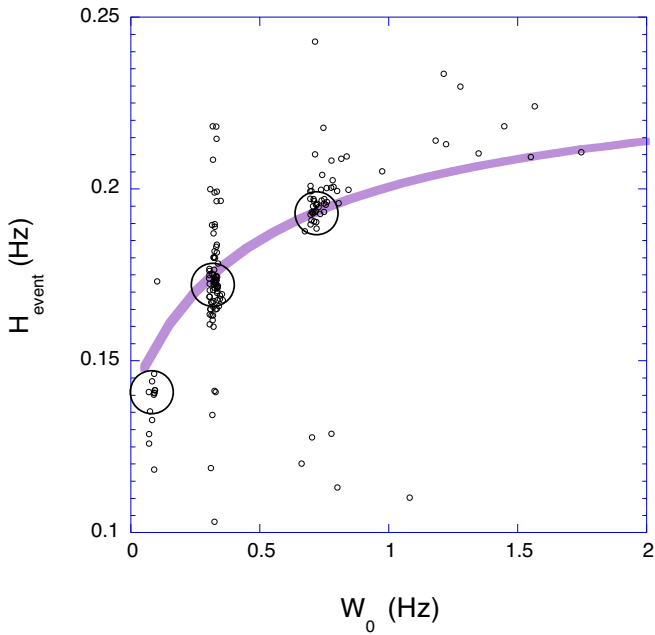}
\caption{Detected events in the energy range of a few hunderd keV in presence of the AmBe neutron source, as a function of $W_{base}$. Note that several points do not actually represent single neutron capture events, but the sites of larger concentration of events at a given  $W_{base}$ (circles) correspond to the neutron peak at that temperature. The continuous line represents the temperature dependence of the expected height $H$ corresponding to the neutron peak as given by (\ref{eq:full-line}).}
\label{fig:neutrons}
\end{center}
\end{figure}

The dependence of the neutron peaks' position on the baseline (i.e.
on temperature) is shown on Fig. \ref{fig:neutrons}. The agreement
of the baseline dependence as given by (\ref{eq:full-line}) is good
except for the very lowest temperatures where an absolute deviation
from (\ref{eq:full-line}) of about 6\,\% is observed at
$W_{base}$=100 mHz. The analysis yields a bolometric heat deposition
of 652$\pm$20 keV at 0 bar, in very good agreement with earlier
measurements, where 650 keV were found under the same conditions \cite{Gren}. Wether
the observed energy deficit of about 110 keV can be significantly
accounted for by topological defects is nevertheless still an open question. An extensive study of the pressure dependence of this
deficit is in progress and might give some new insight on this
question \cite{neutrons-future}.

\section{Scintillation yield in helium of muons and low energy electrons}

The detection of electrons of 7 and 14 keV emitted by a $^{57}$Co
source within one superfluid $^3$He-B bolometer has been reported
recently \cite{manu}. The comparison of the detected electron
spectrum with the bolometric calibration factor of that cell,
determined as described in sections 2-5, yields a deficit of
25$\pm5$\,\% in the bolometrically detected energy. We therefore define the scintillation yield in the following as the fraction of the injected energy that is released as radiation.  

In helium, the
distance between two consecutive collisions of electrons in the keV
range is a fraction of a $\mu$m. This distance is much larger than
that in the case of $\alpha$-particles or the neutron capture
products. Upon electron irradiation, the energy is hence deposited
with much lower density and at no time the density of heat is such that the
superfluid could heat up to the normal state. In the case of
electrons, the energy deficit is therefore entirely due to UV
scintillation. The scintillation yield is very sensitive to the
density of the energy deposition. Our results  can be compared to
measurements by Adams {\em et al.} \cite{adams} who find a 35\,\% UV
scintillation yield in $^4$He upon irradiation by electrons of a few
hundred keV. Nevertheless, as later emphasized by McKinsey {\em et al.}
\cite{mckinsey-pra,mckinsey-nim}, this measurement only covered the
fast emission pulse of the first 10-20\,nanoseconds. A substantial
part of the UV emission, estimated to $\sim$\,50\,\% of the fast
pulse contribution by these authors, takes place at much later times, which
brings the UV scintillation yield to a total of about 50\,\% for high energy
electrons. At lower energies on the keV level, this fraction is
expected to decrease rapidly. Our findings of the scintillation rate
at $\sim$\,10\,keV represent thus a very reasonable low energy
extension of the measurements by McKinsey {\em et al.}

\begin{figure}[ht]
\begin{center}
\includegraphics[scale=0.45,angle=0]{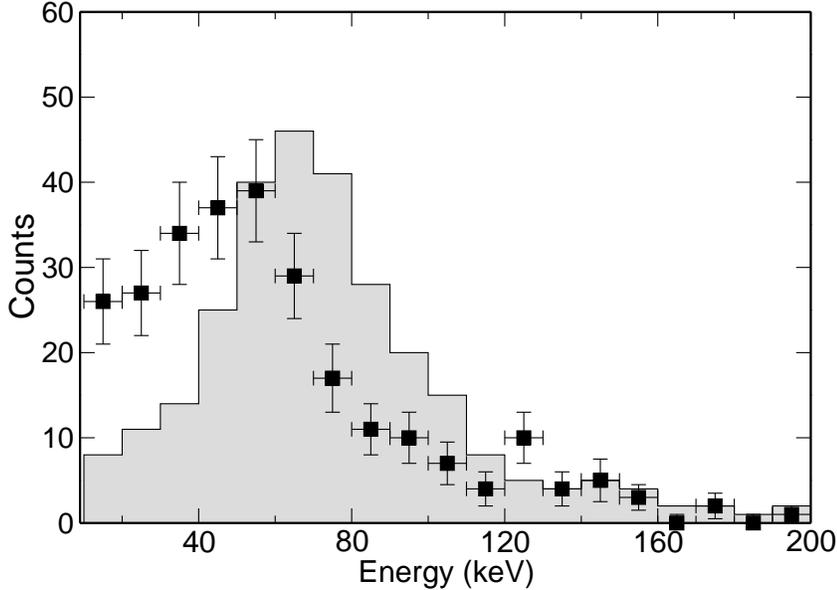}
\caption{Muon spectra: experimental (points) and simulated
(histogram) (data taken from \cite{these-manu}). } \label{fig:muons}
\end{center}
\end{figure}

The muon events also display in our detector a clearly resolved spectrum.  Figure \ref{fig:muons} shows data taken from \cite{these-manu}, comparing numerical simulations to measurements taken within the MACHe3 collaboration on the muon peak. A GEANT 4.0 code computer simulation of the muon energy release in our $^3$He cell, integrated over a large spectrum of incident muon
energies, leads to a broad peak at about 67 keV.  The dispersion of the energy distribution
is a result of the muons tracks crossing the cell geometry in all space directions
with the known angular dependence of radiation intensity \cite{these-manu}. The muon energy release to the target material is rather independent on
the energy of the incident muons in the range from 1 to 10 GeV covering the
maximum of the energy distribution of muons at ground level. 

The
width of the experimentally detected muon spectrum corresponds to the intrinsic resolution
of our detector (3 \% ) combined to the geometrical broadening. The muons produce exclusively electron recoils in
the cell producing delta rays, i.e. electrons scattered from the
atomic shells, energy-distributed following $E_{e}^{-2}$. For one incident muon, roughly 10 of
such electrons have an energy greater than 250 eV and only 5 greater
than 1 keV.

In our experiment we have collected data in the muon energy range
(Fig. \ref{fig:muons}) during 19 hours. A broad maximum of events with an average energy of 52 keV is clearly seen.
The comparison between the bolometric measurement and the simulation of the energy input by muons therfore yields again an
energy deficit of about 25 \%. The missing energy being the result of the scintillation of UV photons as for incident electrons, this observed deficit again allows to quantify the scintillation rate of helium for this type of irradiation.

\begin{table}[ht]
\caption{Measured energy release by different particles and their scintillation rates}
{\footnotesize
\begin{tabular}{@{}lccc@{}}
\hline
particle & {muons} &{electrons} &{$p$+$^3$H} \\ 
\hline
initial kinetic energy (keV) & 2-4 10$^6$ &14  &571+193  \\ 
energy release to target $^3$He (keV)&$\sim$67 &14  &764 \\ 
energy detected bolometrically (keV)&$\sim$52  &11  &652  \\ 
scintillated fraction &0.25 &0.25 &$<0.15$ \\ 
\hline
\end{tabular}\label{table2} }
\end{table}

\section{Conclusions}

We have presented a detailed description of the method of bolometric
calibration of the detector cells based on superfluid $^3$He, by
mechanical pulses. This description provides a general and
comprehensive picture of the bolometer under different conditions of
use (excitation level, temperature).

The results of the calibration are compared to measured heat
depositions by the nuclear neutron capture reactions as well as muon
impacts and low energy electron irradiation. A deficit of about
15\,\% is found in the case of neutrons, in good agreement with
previous measurements at 0 bar \cite{Gren}, in which case this deficit is partly
associated to vortex creation. In the case of high energy muons, as well as electrons in the
10\,keV range, a deficit of about 25\,\% is found which can be entirely
attributed to UV scintillation emission. It is not surprising to find the scintillation rates resulting from these two types of irradiation to be of the same order since the much larger incident energy of cosmic muons is compensated by their larger mass. 

Now that our detector has achieved the required sentitivity for resolving recoils in the keV energy range, we focused on the the feasibility of a electron recoil rejection by parallel scintillation detection. Since we found the UV scintillation yield not to be small for both electrons and muons down to the energy range of interest, this result gives experimental
evidence that the parallel use of a scintillation detector would
allow to reject efficiently electron-, muon- and most likely
$\gamma$-contamination in a bolometric detector based on superfluid
$^3$He for the search of non-baryonic Dark Matter. The optimum design of 
such a parallel scintillation detector will be discussed in a future work.




\begin{thebibliography}{99}

\bibitem{First} G. R. Pickett, in Proc. of the Second European worshop on neutrinos and dark matters detectors, ed. by L. Gonzales-Mestres and D. Perret-Gallix, Frontiers, (1988) 377.
\bibitem{Second} Yu. M. Bunkov, S. N. Fisher, H. Godfrin, A. Gu\'{e}nault, and G. R. Pickett, in Proc. of International Workshop on Superconductivity and Particle Detection, ed. by T. Girard, A. Morales and G. Waysand. World Scientific, (1995) 21-26.
\bibitem{BOLO} D.~I.~Bradley, Yu.~M.~Bunkov, D.~J.~Cousins, M.~P.~Enrico, S.~N.~Fisher, M.~R.~Follows, A.~M.~Gu\'{e}nault, W.~M.~Hayes, G.~R.~Pickett, and T.~Sloan, \PRL{75}{1995}{1887}.
\bibitem{edelweiss}V. Sanglard {\it et al.}, \PRD{71}{2005}{122002}.
\bibitem{cdms} D.S. Akerib {\it et al.},  \PRL{96}{2006}{011302}.
\bibitem{machetnim}F.~Mayet {\it et al.}, \NIMA{455}{2000}{554}.
\bibitem{machetplb}F.~Mayet {\it et al.}, \PLB{538}{2002}{257}.
\bibitem{Gren} C.~B\"{a}uerle, Yu.~M.~Bunkov,  S.~N.~Fisher, H.~Godfrin, and G.~R.~Pickett,
{Nature} {\bf 382} (1996) 332.
\bibitem{nature85} W. Zurek, {Nature} {\bf 317}, (1985) 505.
\bibitem{Gren2} C.~B\"{a}uerle, Yu.~M.~Bunkov,  S.~N.~Fisher, and H.~Godfrin, { Phys.  Rev.} {\bf B57} (1998) 14381.
\bibitem{moriond} C.B. Winkelmann,  E. Moulin, Yu.M. Bunkov, J. Genevey, H. Godfrin,
J. Macias-P\'erez, J.A. Pinston, and D. Santos.
"MACHE3, a prototype for non-baryonic dark matter search: KeV event
detection and multicell correlation." In "Exporing the Universe,
XXXIX Rencontres de Moriond", ed. Giraud-Heraud and Thanh Van, The
Gioi pbl., (2004) 71.
\bibitem{manu}  E. Moulin, C. Winkelmann, J.F. Macias-P\'erez, Yu.M. Bunkov, H. Godfrin,
D. Santos, \NIMA{548}{2005}{411}.
\bibitem{fisher} S. N. Fisher, A. M. Gu\'{e}nault, C. J. Kennedy, and G. R. Pickett, {Phys. Rev. Lett.}  {\bf 63} (1989) 2566.
\bibitem{thermo} C. B. Winkelmann, E. Collin, Yu. M. Bunkov, and H. Godfrin, {J. Low Temp. Phys.} {\bf 135} (2004) 3.
\bibitem{fisher-bis} S. N. Fisher, G. R. Pickett, and  R. J. Watts-Tobin, {J. Low Temp. Phys.}  {\bf 83} (1991) 225.

\bibitem{bradley} D. I. Bradley, {Phys. Rev. Lett.}  {\bf 84} (2000) 1252.
\bibitem{vollhardt} D. Vollhardt and P. W\"{o}lfle, {\it The Superfluid Phases of Helium 3} \\
Taylor \& Francis (1990).

\bibitem{meyer} J. S. Meyer and T. Sloan, {J. Low Temp. Phys.}  {\bf 108} (1997) 345.
\bibitem{mckinsey-pra} D. N. McKinsey {\em et al.}, {Phys. Rev.}{\bf A67} (2003) 62716.
\bibitem{neutrons-future} J. Elbs, C. B. Winkelmann, E. Collin, Yu. M. Bunkov, and H. Godfrin, in progress.
\bibitem{adams} J. S. Adams, Y. H. Kim, R. E. Lanou, H. J. Maris, and G. M. Seidel, {J. Low Temp. Phys.}  {\bf 113} (1998) 1121.
\bibitem{mckinsey-nim} D. N. McKinsey {\em et al.}, {Nucl. Instr. and Meth.} {\bf A516} (2004) 475.
\bibitem{these-manu} E. Moulin, Ph.D. Thesis, Universit\'e Joseph Fourier, 2005, unpublished.


\end{thebibliography}
\end{document}